\documentclass[doublecol]{epl2} 
\usepackage{epstopdf}
\usepackage{graphicx}
\usepackage{dcolumn}
\usepackage{bm}
\usepackage{hyperref}
\usepackage{subfigure}
\hypersetup{colorlinks=true, citecolor=blue, urlcolor=blue, linkcolor=blue}
\usepackage{amssymb}
\usepackage{color}
\usepackage{graphicx}
\usepackage{amsmath}
\usepackage{mathrsfs}
\usepackage{times}
\usepackage{subeqnarray}
\usepackage{cases}
\usepackage{bm}
\usepackage{enumitem}
\usepackage{booktabs}
\usepackage[table,xcdraw]{xcolor}

\title{Emergent route towards cooperation in interacting games: the dynamical reciprocity}
\shorttitle{Emergent route towards cooperation in interacting games: the dynamical reciprocity}

\author{Qin-Qin Wang\inst{1} \and Ri-Zhou Liang\inst{1} \and Ji-Qiang Zhang\inst{2,3} \and Guo-Zhong Zheng\inst{1} \and Lin Ma\inst{1} \and Li Chen\inst{1,4}\thanks{E-mail:chenl@snnu.edu.cn}}
\shortauthor{Qin-Qin Wang \etal}

\institute{
  \inst{1} School of Physics and Information Technology, Shaanxi Normal University, Xi'an 710062, China\\
  \inst{2} School of Physics and Electronic-Electrical Engineering, Ningxia University, Yinchuan 750021, P. R. China\\
 \inst{3} Beijing Advanced Innovation Center for Big Data and Brain Computing, Beihang University, Beijing 100191, P. R. China\\
  \inst{4} Robert Koch-Institute, Nordufer 20, 13353 Berlin, Germany
}

\abstract{
The success of modern civilization is built upon widespread cooperation in human society, deciphering the mechanisms behind has being a major goal for centuries. A crucial fact is, however, largely missing in most prior studies that games in the real world are typically played simultaneously and interactively rather than separately as assumed. Here we introduce the idea of interacting games that different games coevolve and influence each other's decision-making. We show that as the game-game interaction becomes important, the cooperation phase transition dramatically improves, a fairly high level of cooperation is reached for all involved games when interaction goes to be strong. A mean-field theory indicates that a new mechanism --- \emph{the dynamical reciprocity}, as a counterpart to the well-known network reciprocity, is at work to foster cooperation, which is confirmed by the detailed analysis. This revealed reciprocity is robust against variations in the game type, the population structure, and the updating rules etc, and more games generally yield a higher level of cooperation. Our findings point out the great potential towards high cooperation for many issues are interwoven with each other in the real world, and also the possibility of sustaining decent cooperation even in extremely adverse circumstances.
}

\begin{document}

\maketitle

\section{Introduction}
Recent withdrawals of the United States from a couple of ``groups'' like WHO, Paris Agreement, UNESCO etc. signifies a degraded cooperation at the global scale. Any solution to this sort of problems requires an understanding of what processes drive and maintain human cooperation and what measures or institutions could be implemented for its promotion. The key question to be addressed is: why entities help each other who could potentially be in competition and incur a cost to themselves? As the paradigm of \emph{homo economicus} shows, people always try to maximize their earnings and avoid irrational investments, which inevitably leads to the tragedy of the commons \cite{hardin1968tragedy}.

Important progresses have been made with the help of evolutionary game theory~\cite{Nowak2004Evolutionary} by analysing the stylized social dilemmas such as prisoner's dilemma and the public goods game.
Several mechanisms are proposed \cite{Nowak2006Five} in the past several decades, such as reward and punishment \cite{Sigmund2001Reward}, social diversity~\cite{Santos2008Social}, direct~\cite{trivers1971evolution} or indirect reciprocity~\cite{nowak1998evolution}, kin~\cite{hamilton1964genetical} or group selection \cite{keller1999levels,Queller1964Group}, spatial or network reciprocity \cite{nowak1992evolutionary}. In particular, theoretically accounting for the fact that human populations are highly organized and individuals interact repeatedly with their immediate neighbors can support cooperation \cite{nowak1992evolutionary}.  The rationale behind is that a structured neighborhood facilitates the formation of cooperator clusters, which effectively resist the invasion of defectors, as opposed to the well-mixed scenario. The ensuing years have witnessed a wealth of theoretical studies that further confirm this so-called network reciprocity for various population structures~\cite{szabo2007evolutionary}. However, recent human behavioral experiments show that structured populations do not promote cooperation in general \cite{Gracial2012Human,Carlos2012Heterogeneous}, at least some conditions combining the game parameters and the population structure must be met for cooperation to thrive \cite{Rand2014Static}. One explanation is that the complexities of human psychology make humans switch strategies frequently that the assortment fails in the static networks \cite{traulsen2010human}. But dynamic networks indeed offer an escape because the players are allowed to adjust social ties and they are more like to cooperate under this peer pressure \cite{Rand2011Dynamic, Katrin2011Co}. This unsatisfactory situation implies that some essential elements could be missing in current game-theoretic models and the experiment-driven modeling approach is needed. Note that, in most of these studies a single game is considered, and they focus on the factors of interest like the underlying structures of population, or the impact of some dynamical processes, e.g. the punishment or reward; the conclusions drawn are supposed to be applicable in more general circumstances.

Games, however, may not be unfolded in isolation but often in parallel. For instance, we humans are engaged in different activities, works, sports, and recreations; colleagues in a company could work on a couple of concurrent projects; and countries have to deal with a whole range of conflicts such as trade war, security issues, diplomatic crisis etc. Only when the evolution of these games is independent from each other, the modeling efforts based on a single game are then reasonable as most of existing work assumed. Observations in aforementioned contexts, however, suggest that the decision-making of entities in one game is often conditioned by what happened in another. Similar observations are also made in biological games, like chimpanzees are more likely to groom their fellows if they are skillful in hunting, and vice versa, and actually these two behaviors together with sharing food, joint patrol the borders, support one another in conflicts etc are all correlated in the chimpanzee's social life~\cite{hammerstein2003genetic}.

A closely related research line is the multigame dynamics, the existing work shows that dynamical inconsistencies are already possible when two or more non-repeated games are coupled~\cite{cressman2000evolutionary,chamberland2000example,hashimoto2006unpredictability,venkateswaran2019evolutionary}, meaning that the eventual fate of games cannot be inferred from the single game dynamics. A more recent work starts to study the repeated scenario and an evolutionary framework of the so-called multichannel games is proposed~\cite{donahue2020evolving}, where they find that the fixed game linkage is able to enhance cooperation in all games engaged in general. Still, fundamental questions remain: \emph{what typical evolutionary dynamics are expected when more games are engaged, to what extent would such game-game interaction alter the classic cooperation mode of single game, and any new cooperation scenario arises therein?}

In this Letter, we mainly study two symmetrically interacting games, where they have a stake in each other, and focus on clarifying the impact of game-game interaction. We reveal a new type of reciprocity rooted in the game-game interaction that is able to maintain high levels of cooperation. In particular, fairly high cooperation is expected when the interaction goes to the extreme that the decision-making of a given game is completely conditioned by the other and vice versa. The mechanism behind lies in the new types of interactions that lead to a persistent advantage of cooperators. Furthermore, the uncovered reciprocity is found to be quite robust and more games generally lead to be more cooperative for all games. For a detailed account of this study, see~\cite{Liang2021arXiv}.

\begin{figure}
\centering
\includegraphics[width=0.8\linewidth]{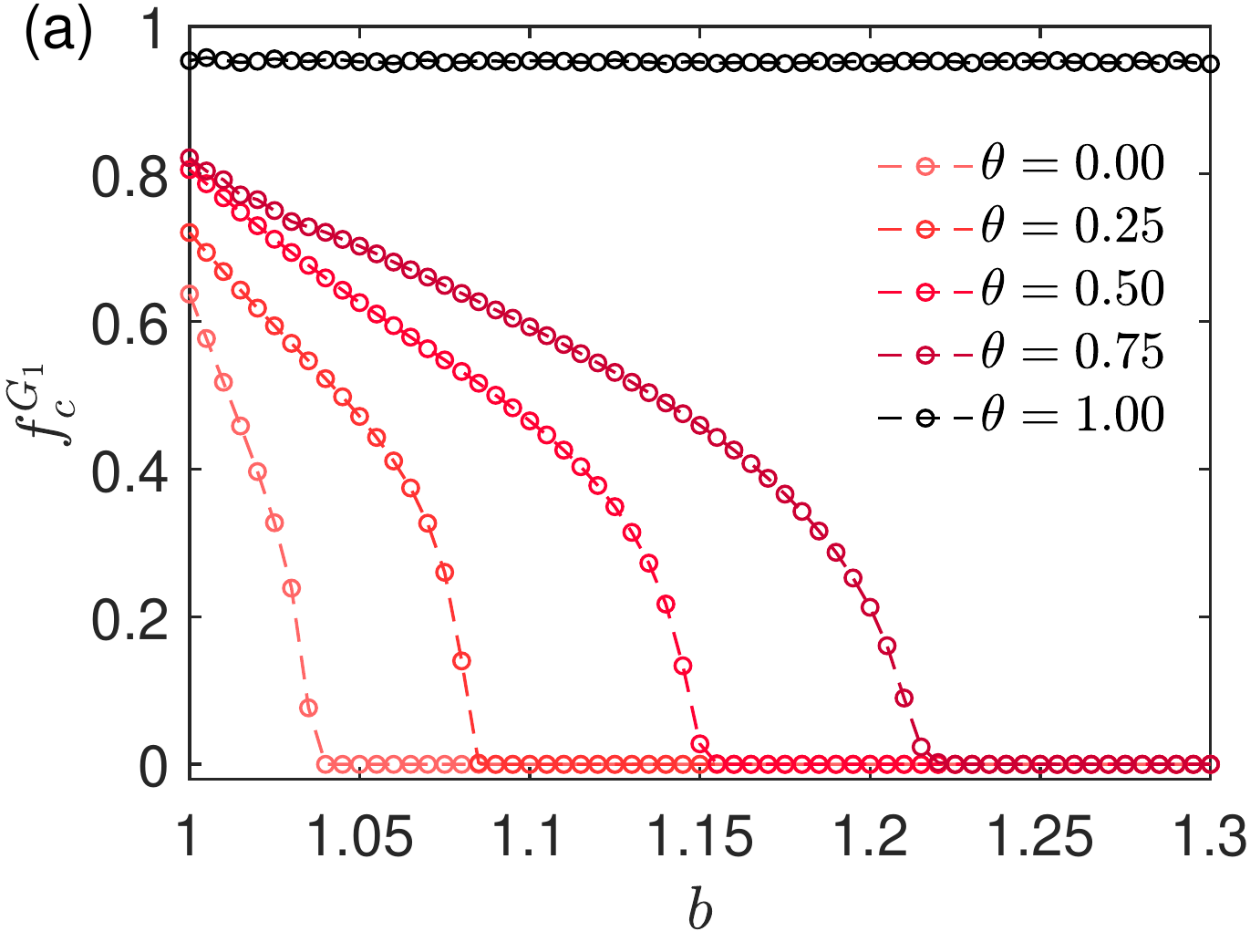}
\includegraphics[width=0.8\linewidth]{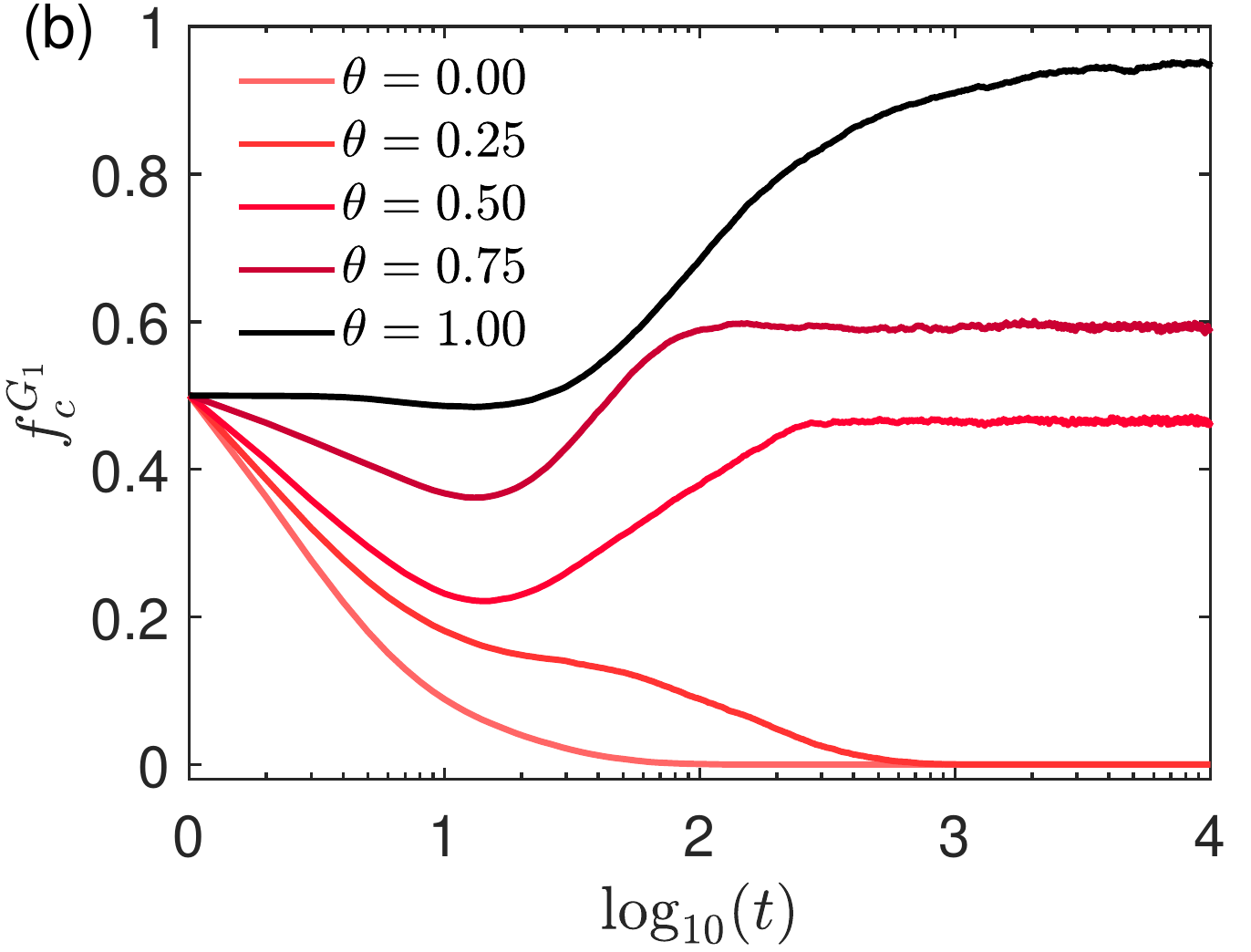}
\caption{
The cooperation evolution of two symmetrically interacting PD on the 2d square lattice.
(a) Phase transitions of cooperation prevalence regarding game $G_1$ ($f^{G_1}_C=f_{CC}+f_{CD}$) versus the temptation $b$ is shown for interaction strengths $\theta=0,0.25,0.5, 0.75$ and 1. Due to the symmetry, $f^{G_2}_C\approx f^{G_1}_C$ (data not shown for visual clarity).
(b) Typical time series are shown with fixed $b=1.1$. Parameters: $L=1024$ and $K=0.1$, the random initial condition for both games, data is obtained over 50 ensemble averages after transient in (a).
}
\label{fig:PT}
\end{figure}

\section{Modeling two interacting games}
Suppose that two games $\mathbb{G}\!=\!\{G_1, G_2\}$ are played simultaneously in a population composed of $N$ players, where they are located on an $L\times L$ square lattice with a periodic boundary condition. They can adopt one of the two strategies for each game: cooperation ($C$) or defection ($D$), i.e. $\mathbb{S}_1=\{C,D\}$. Therefore, there are four possible states $\mathbb{S}_2=\{XY|CC, CD, DC, DD\}$ in the two interacting games, where $X,Y$ represent the state regarding game $G_{1,2}$ respectively. For simplicity, we resort to the pairwise game defined as follows: mutual cooperation brings both a reward $R$, mutual defection leads to a punishment $P$ for each, and mixed encounter yields the cooperator a sucker's payoff $S$ yet a temptation $T$ for the defector. Their ranking determines the game type.
Here, we follow the common practice for a weak prisoner's dilemma (PD) with $R=1$, $P=S=0$, $T=b>1$ for both games if not stated otherwise.

Following the standard Monte Carlo (MC) simulation procedure, firstly a game $g\in \mathbb{G}$ is chosen at random to play in an elementary step, a player $i$ is then randomly chosen and accumulates its payoff $\Pi_i$. Next, one of $i$'s neighbors $j$ is picked randomly, and acquires its payoff $\Pi_j$ as well. Lastly, player $i$ adopts $j$'s strategy regarding game $g$ with a probability according to the Fermi rule~\cite{szabo1998evolutionary}
\begin{equation}
W^g_{j\rightarrow i}=\frac{1}{1+\exp[(\widehat{\Pi}_i^g-\widehat{\Pi}_j^g)/K]} ,
\label{model1}
\end{equation}
\begin{equation}
\widehat{\Pi}_{i,j}^{G_{1,2}}=(1-\theta)\Pi_{i,j}^{G_{1,2}}+\theta \Pi_{i,j}^{G_{2,1}},
\label{model2}
\end{equation}
where $\widehat{\Pi}_{i,j}^g$ is \emph{the effective payoffs}, which captures the reality that to imitate, players compare the overall payoff profiles in all games rather than simply the one under play. Therefore the decision is made based upon a combination of both payoffs, a simple case is as shown in Eq. (\ref{model2}). We interpret the weight $\theta\in[0,1]$ as the game interaction strength, a larger value means a stronger impact of the other game; two extreme cases $\theta=0,1$ correspond to the two independent games and the cross-playing scenario, respectively. $K$ is a temperature-like parameter, measuring the uncertainties in the imitation process, its inverse can be interpreted as the selection intensity in biology or the bounded rationality in economical contexts. A full MC step consists of $2\!\times\!L\!\times\!L$ elementary steps, where every player is updated once for each game on average. Simulations are carried out for $L=1024$, and the data for the cooperator fractions are averaged over $10^6$ MC steps after a transient period of $10^6$ steps. 

\section{Results}
Varying the game interaction strength $\theta$, we observe a continuing promotion in the cooperation prevalence $f_c$ as a function of the temptation $b$ [see Fig.~\ref{fig:PT}(a)]. For the independent game case where $\theta=0$, a second-order phase transition (PT) for cooperation is seen but the cooperation region is rather small with the critical temptation $b_c\approx1.038$, beyond which the cooperators become extinct. As $\theta$ is increased, $b_c$ is shifted to the right, the prevalence also becomes higher, the cooperation is lifted. Finally, as $\theta\rightarrow 1$, this promotion is maximal, where the PTs become absent and nearly full cooperation is seen across the whole parameter region $1\leq b \leq2$ for both games. This is quite unexpected since in the cross-playing scenario ($\theta=1$), the decision-making of a game is entirely blind to its own payoff. This observation of promotion is strengthened by the corresponding time series by fixing $b=1.1$ shown in Fig.~\ref{fig:PT}(b), where the initial decrease in $f_c$ is also inhibited when $\theta$ becomes large.

\begin{figure}
\centering
\includegraphics[width=1\linewidth]{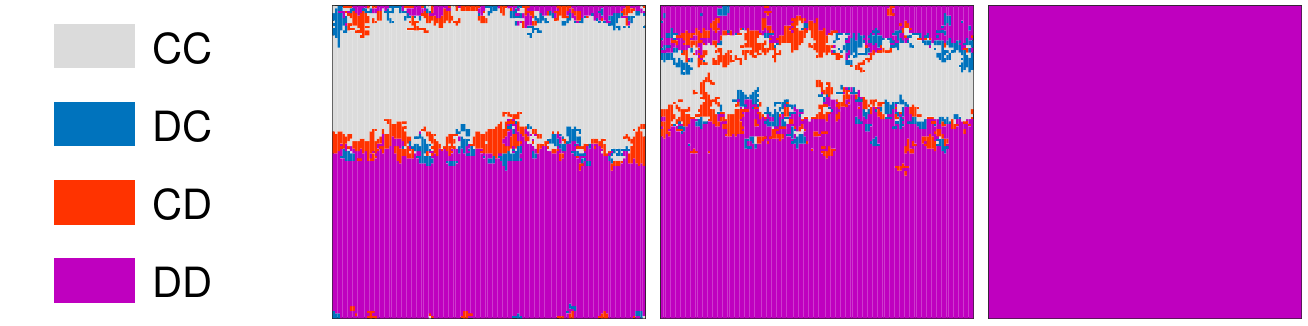}
\includegraphics[width=1\linewidth]{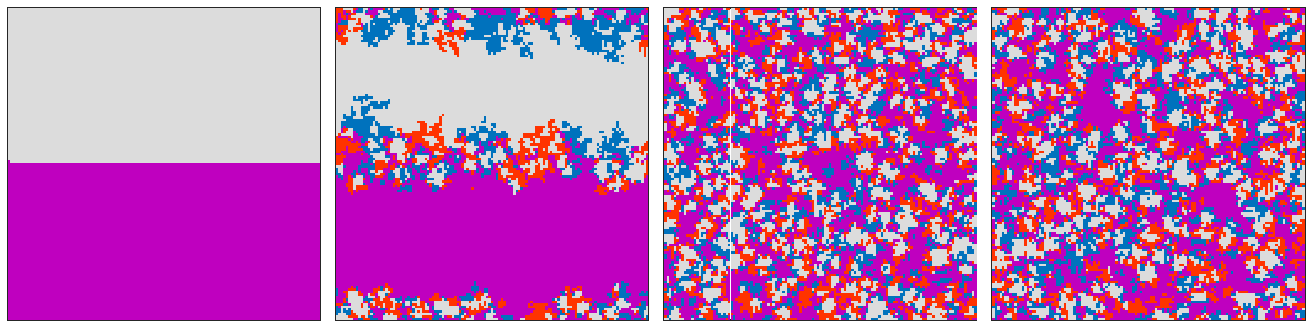}
\includegraphics[width=1\linewidth]{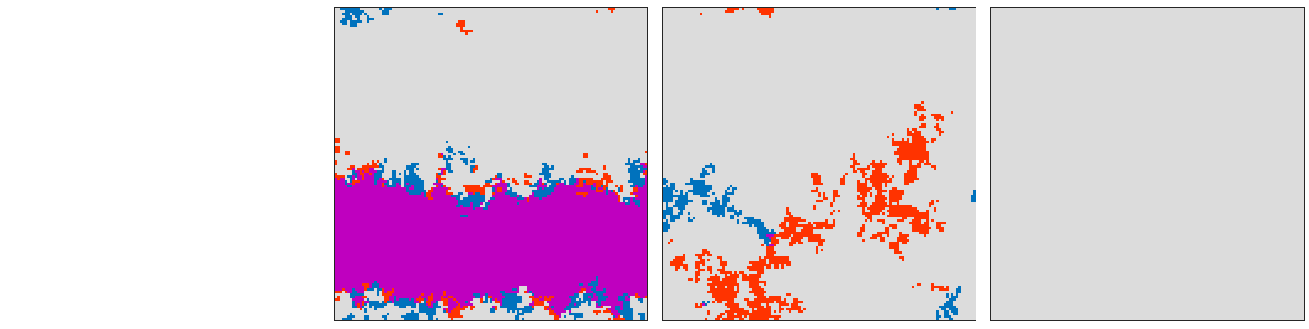}
\caption{
The evolution of cooperation patterns for $\theta=0$ (top row), 0.5 (middle row), and 1(bottom row). The system is prepared with full cooperators CC within upper half domain versus full defectors DD within another half (the leftmost panel). The characteristic snapshots are taken at $t=0,50,100,200,300$ for $\theta=0$, whereas $t=0,100,1000,5000,10000$ for the others.
Parameters: the domain is of size $128\times128$, $K=0.1$, and $b=1.1$.
}
\label{fig:pattern}
\end{figure}

To gain some intuition of how the game-game interaction affects cooperation, we first look at how the spatiotemporal evolution is influenced. Fig.~\ref{fig:pattern} shows the cases of $\theta=0,0.5,1$, but starting from a bulk initial condition because it is more intuitive without altering $f_c$ compared to the random initial condition cases. When $\theta=0$, defectors dominate in both games, $DD$ players invade the $CC$'s domain, and cooperators quickly go extinct.  At the intermediate strength $\theta=0.5$, this advantage disappears where all four species coexist. In the other extreme $\theta=1$, a reversed invasion is seen where $CC$ players dominate and take over the whole domain in the end. This suggests that a reversed advantage is expected between cooperators and defectors as the game-game interaction is engaged.

\section{A mean-field theory}
To understand the rationale behind, we develop a mean-field theory based on the replicator equation~\cite{taylor1978evolutionary,roca2009evolutionary}, where the evolution of the four fractions with respect to each game depends on their relative fitness measured by the payoffs that can be formally described as $\dot{f}_s=f_s(\widehat{\Pi}_s-\bar{\Pi})$, where $g\!\in\! \mathbb{G}$, $s\!\in\! \mathbb{S}_2$, and $\bar{\Pi}=\sum_s f_s\widehat{\Pi}_s$ is the average fitness. With some algebra (see~\cite{Liang2021arXiv}), we obtain the ordinary differential equations of cooperator fraction for game $G_1$ (i.e. $f^{G_1}_C=f_{CC}+f_{CD}$, the exchange of 1 and 2 applies for game $G_2$) as \\
\small
\begin{equation}
\dot{f}^{G_1}_C = f^{G_1}_C f^{G_1}_D (\Pi^{G_1}_C-\Pi^{G_1}_D)+(f_{CC}f_{DD}-f_{CD}f_{DC}) (\Pi^{G_2}_C-\Pi^{G_2}_D),
\label{Eq:meanfield}
\end{equation}
\normalsize
where $\Pi^{G_{1,2}}_{C,D}$ are the fitness in game $G_1$ or $G_2$ respectively for the cooperators and defectors. The first term in the rhs. is well-known in the single game scenario~\cite{smith1982evolution} that comes from the game under play, meaning that the fitness advantage in cooperators $\Pi^{G_1}_C>\Pi^{G_1}_D$ converts the defectors into cooperators when they meet up. The second term is new that captures the game-game interaction. Specifically, the impact of the other game is through two interacting pairs: i) when $CC$ players come across $DD$, the advantage of cooperators in game $G_2$ ($\Pi^{G_2}_C>\Pi^{G_2}_D$) also facilitates the proliferation of cooperators in game $G_1$ due to the game-game correlation; ii) unexpectedly, in the opposite case when $\Pi^{G_2}_D>\Pi^{G_2}_C$, the advantage of defectors in $G_2$ also helps the growth of cooperators in $G_1$ when $CD$ encounters $DC$ players. Therefore, the above analysis shows that potentially there are now new dynamical routes at work towards cooperation in addition to the one in the single game case.

\begin{table*}[btp]
\begin{center}
\begin{tabular}{@{}rlll@{}}
\hline\hline
\toprule
\multicolumn{1}{l}{} & \multicolumn{1}{c}{~~~~~~~~~~Individual scenario~~~~~~~~~~}                                                              & \multicolumn{1}{c}{~~~~~~~~~~~~~Bulk scenario~~~~~~~~~~~~~}                                                                    &  \\ \midrule
\hline
Invasion~~        & \begin{tabular}[c]{@{}l@{}}
                                  \cellcolor[HTML]{EFEFEF}CC + DD $\xrightarrow{  G_1/G_2  }$ DC/CD + DD\\ \end{tabular}
                                 & \begin{tabular}[c]{@{}l@{}}
                                  \cellcolor[HTML]{EFEFEF}CC + DD $\xrightarrow { G_1/G_2 }$ CC + CD/DC\end{tabular}                                                                       &\\
Neutral~~         & \begin{tabular}[c]{@{}l@{}} 
                                    CC + DC $\xrightarrow{    G_1    }$ 2CC or 2DC\\
                                    CC + CD $\xrightarrow{    G_2    }$ 2CC or 2CD\\
                                    DD + CD $\xrightarrow{    G_1    }$ 2DD or 2CD\\
                                    DD + DC $\xrightarrow{    G_2    }$ 2DD or 2CD\\\end{tabular}
                                    & \begin{tabular}[c]{@{}l@{}} 
                                    CC + DC $\xrightarrow{    G_1    }$ 2CC or 2DC\\
                                    CC + CD $\xrightarrow{    G_2    }$ 2CC or 2CD\\
                                    DD + CD $\xrightarrow{    G_1    }$ 2DD or 2CD\\
                                    DD + DC $\xrightarrow{    G_2    }$ 2DD or 2CD\\\end{tabular}  &\\
Catalyzed~~      & \begin{tabular}[c]{@{}l@{}}
                                   \cellcolor[HTML]{EFEFEF}CD + DC $\xrightarrow{   G_1/G_2   }$ CD/DC + CC\\ \end{tabular}
                                  & \begin{tabular}[c]{@{}l@{}}
                                   \cellcolor[HTML]{EFEFEF}CD + DC $\xrightarrow{   G_1/G_2   }$ DC/CD + DD\end{tabular}                                                                       &  \\ \bottomrule
\hline
\end{tabular}
\end{center}
\caption{Classification of interactions in two interacting PD games for the cross-playing scenario ($\theta=1$), where all six pairwise interactions can be classified into three categories, in either individual or bulk scenario. In invasion interactions, the payoff advantage of a given strategy in either game is explicitly transformed into its reproduction. For neutral type, the state change is purely random, no net conversion towards cooperation or defection is expected. In the last category, the advantage of defectors/cooperators regarding a given game cross-catalyzes the production of cooperators/defectors in the other game. The last two categories are only possible in interacting games.}
\label{tab:2games}
\end{table*}

\begin{figure}[tbp]
\centering
\includegraphics[width=1.0\linewidth]{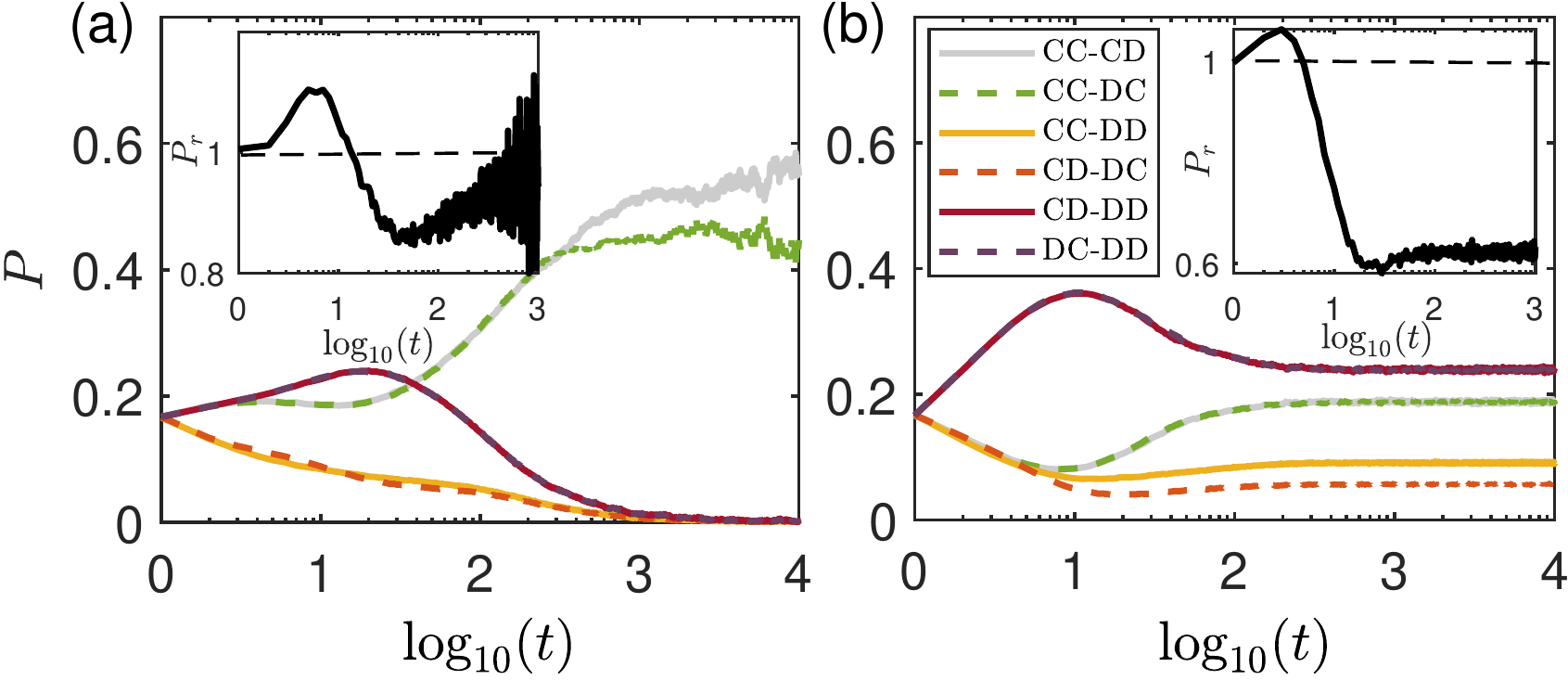}
\caption{
Time evolution of all six interface proportions for two interacting PD games starting with random initial conditions for $\theta=1$ (a) and $0.5$ (b), respectively. Inset show the relative fraction of the two non-neutral types $P_r=P_{_{CD-DC}}/P_{_{CC-DD}}$.
Parameters: $b=1.1$ and $L=1024$ for the 2d square lattice.
}
\label{fig:interface}
\end{figure}

\section{Mechanism analysis}
To be more specific, it's helpful to look into all six interactions in details in our lattice system, as listed in Table~\ref{tab:2games}.
Here, we distinguish two scenarios --- individual and bulk scenarios. In the former, we only focus on the evolution of the interaction pairs when without knowledge of their surroundings such as the random state configuration. The bulk scenarios apply for the circumstance when players of the same type are well-bulked, both intra- and inter-bulk play are incorporated. The two scenarios are typically present in the early phase of evolution and afterwards, respectively. For simplicity, we consider the cross-playing case, where the six pairs of interactions can be classified into three categories: invasion, neutral, and catalyzed type for both scenarios (Table~\ref{tab:2games}). While the neutral type of interactions brings no net effect on cooperation, the other two categories determine the evolution of cooperation prevalence, though they always have the opposite effects either in individual or bulk scenario.

Typical evolution of all interactions is shown in Fig.~\ref{fig:interface}(a) starting from random initial conditions, where the evolution can be roughly divided into two stages. (i) At the early stage $t\!<\!t_c$ ($t_c\!\approx\!10$ MC steps) when no clear clusters are formed and thus the individual scenario applies, the proportion $P_{_{CD-DC}}>P_{_{CC-DD}}$ is detected, meaning that catalyzed interactions dominate over the invasion ones; and according to the evolutionary dynamics in Table~\ref{tab:2games}, a net production of cooperation is expected. (ii) When $t>t_c$, clusters are gradually formed, where both the size and compactness increase (see~\cite{Liang2021arXiv}), therefore the bulk scenario sets in. Interestingly, a crossover is seen that $P_{_{CD-DC}}<P_{_{CC-DD}}$, the reversed dominance again yields a net increase of cooperators since $CC\!-\!DD$ pairs in bulk scenario favor the cooperators (Table~\ref{tab:2games}). Therefore, cooperation is preferred in the whole evolutionary processes.
Back to the mean-field equation Eq.(\ref{Eq:meanfield}), our analysis indicates that the second term always brings a positive contribution to the cooperation evolution. Thereinafter, we term this mechanism caused by game-game interactions as the \emph{dynamical reciprocity}.
It also works for cases with $\theta<1$, as shown in Fig.~\ref{fig:interface}(b). However, the dynamical reciprocity only works in structured population, no promotion is seen in the well-mixed population (see~\cite{Liang2021arXiv}). 

\section{Robustness} Within structured population, the revealed reciprocity is quite robust. In~\cite{Liang2021arXiv}, we show that when the interacting game is extended to be general pairwise games (including snowdrift game and stage hunt etc), a similar cooperation promotion is still observed irrespective of the game type, and fairly high cooperation is expected for the whole parameter domain when games are cross played. We also show it is also applicable to a multiplayer game (the public goods game); Robust observations are made in model variants such as asymmetrically interacting games, games with different updating rules (like replicator rule, Moran rule, follow-the-best rule etc~\cite{roca2009evolutionary}), with different time-scales, and even with two different games, i.e. a PD is coupled with a snowdrift game.

Additional structural complexities~\cite{Liang2021arXiv} from underlying populations like small-world networks and Erd\H{o}s-R\'enyi random topologies also do not change the working of the reciprocity. The structural heterogeneity neither alters the promotion trend, as shown in the case of scale-free networks.

\begin{figure}[htbp]
\centering
\includegraphics[width=1.0\linewidth]{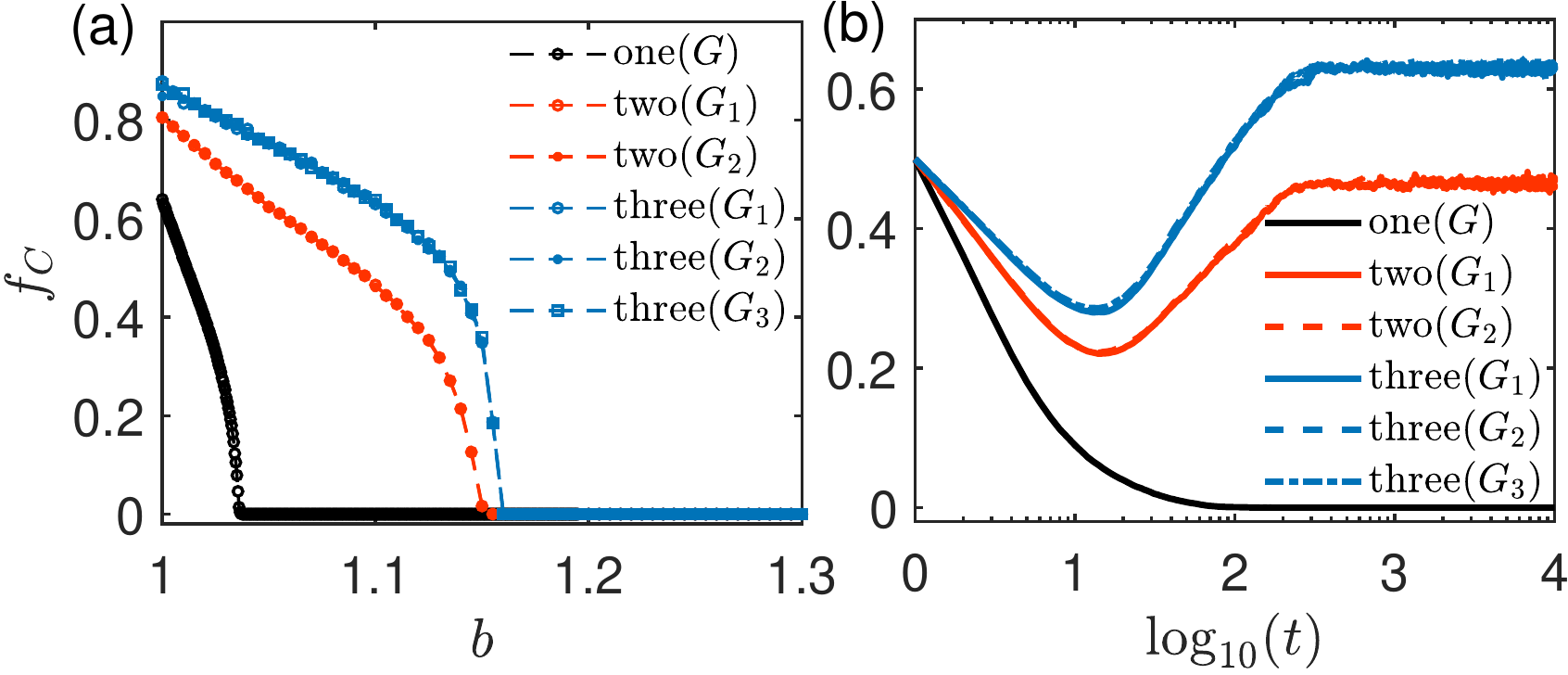}
\caption{
(a) Phase transitions of cooperation prevalence  for one-, two-, and three-game cases, where the each game is of equal contribution in the effective payoffs for the later two cases (i.e. $\theta=1/2$ and $1/3$, respectively). (b) Time series for fixed $b=1.1$.
Parameters: $L=1024$ for the 2d square lattice and $K=0.1$.
}
\label{fig:more}
\end{figure}

In particular, when the number of engaged games increases, a higher level of cooperation is expected in general, see Fig.~\ref{fig:more}, where equal contribution for each game is assumed when more than one game is played simultaneously. Since potentially there are many issues interwoven with each other in reality, much higher cooperation is expected than the case when only a single game is unfolded.

\section{Conclusions}
In summary, the discussed game-game interaction is a natural ingredient that may underpin a wealth of issues, from complex behaviors in animals, to inter-personal activities in daily life, and even to international relationships at the global scale. The potential for being highly cooperative, as revealed here points out a promising route towards a cooperative world. It is worthwhile to emphasize that contrary to the network reciprocity, where the underlying structure of population plays the key role~\cite{nowak1992evolutionary,szabo2007evolutionary}, including the interdependent network reciprocity~\cite{Wang2013interdependent,jin2014spontaneous,xia2015heterogeneous}. The mechanism behind the promotion here stems instead from the dynamical interaction among different games. Our results suggest that the dynamical reciprocity could constitute a new category of mechanisms behind the emergence of cooperation.

On the theoretic side, our finding of ``more is different''~\cite{anderson1972more} calls for more systematic investigations in specific contexts, since the revealed mechanism may offer valuable inspiration to avoid the cooperation crises. On the experimental side, behavioral experiments are needed to justice the dynamical reciprocity in realities and unveil other complexities that may arise in interacting games.

\acknowledgments
This work is supported by the National Natural Science Foundation of China under Grants 61703257 and 12075144, and by the Fundamental Research Funds for the Central Universities GK201903012. L. C. acknowledges the enlightening discussions with Dirk Brockmann (HU and RKI) in the early phase of the project and Ying-Cheng Lai (ASU) for helpful comments.

\end{document}